\documentclass[10pt,letterpaper,twocolumn]{article}
\usepackage{ol2}
\usepackage{graphicx}
\usepackage[latin1]{inputenc}
\usepackage{amssymb}
\usepackage[normalem]{ulem}

\begin{document}
\twocolumn[
\title{A thermal bistability-based  method  for real-time optimization \\ of ultra-low threshold whispering gallery mode microlasers}
\author{Guoping Lin$^{1,2}$, Y. Candela$^1$, O. Tillement$^3$, Zhiping Cai$^2$, V. Lefèvre-Seguin$^1$ and J. Hare$^1$ }
\address{
$^1$Laboratoire Kaster Brossel, ENS, UPMC, CNRS -- 24 rue Lhomond, 75231 PARIS cedex 05, France\\
$^2$ Department of Physics, Xiamen University, Xiamen 361005, Fujian, P. R. China\\
$^3$ LPCML, Université Claude Bernard Lyon 1, 10 rue André-Marie Ampère,  69622 Villeurbanne cedex,  France\\
$^*$Corresponding author: \texttt{Jean.Hare@lkb.ens.fr}
}


\begin{abstract}
A method based on thermal bistability for ultra-low threshold microlaser optimization is demonstrated. When sweeping the pump laser frequency across
a pump resonance,  the dynamic thermal bistability slows down the power variation. The resulting
lineshape modification enables a real-time monitoring of the  laser characteristic. We demonstrate this method for a functionalized microsphere
exhibiting a sub-microwatt laser threshold. This approach is confirmed by comparing the results with a
step-by-step recording in quasi-static thermal conditions.
\end{abstract}
\ocis{
140.3945,   
140.3410,   
140.3530,   
140.6810,   
130.3990    
} 
]

Optical microcavities have drawn a large interest in the last two decades and received numerous applications, including Cavity-QED and photonic
devices, like light emitting diodes or microlasers. A special attention has been devoted to whispering gallery mode (WGM) microcavities induced by
surface tension, like microspheres and microtoroids. They benefit from the sub-nanometer roughness of molten silica, which results in very high
quality factors $Q$, enabling laser operation with a pump power in the sub-microwatt range \cite{SandoghdarTreussart1996}. These ultra-low-threshold
lasers have been obtained by silica functionalization with embedded rare earth ions. Different doping techniques, like fiber doping, sol-gel
coating, ion implantation, co-deposition or co-sputtering have been successfully used with Nd$^{3+}$, Er$^{3+}$, and Yb$^{3+}$ or
mixtures\cite{SandoghdarTreussart1996,LissillourFeron2000,YangArmani2003,OstbyYang2007}.

In this paper, we report on a new method for the fast measurement of the laser light-light characteristic based on thermal effect, and allowing real-time optimization. In monolithic microcavities,  self-heating 
by the minute dissipated optical power induces a negative frequency shift, resulting in a bistable behavior when this shift exceeds the cold cavity linewidth 
\cite{BraginskyGorodetski1989,RokhsariSpillane2004,CarmonYang2004}.
For decreasing laser frequency, the heating pushes cavity resonance frequency in the same direction, and they remain close to resonance, as long as the thermal shift compensates the laser offset. The loss of the 
resonance results in a rapid drop of the coupling. In reverse direction, the thermal shift and the laser scan have opposite directions, and the line is narrowed. Due to the high-$Q$ and  the small heat capacity 
of the cavity, the bistability threshold is very low, typically in the microwatt range or less.  At low scanning rate (typically up to $100\,\mathrm{Hz}$) the heating affects the cavity as a whole, and only one 
cooling time is involved, while for increasing scanning rate a dynamical thermal effect appears which depends on the heat diffusion details.  
Our method exploits the low frequency regime, with a decreasing pump frequency so that the resonance shifts in the same direction as the laser.

\begin{figure}
\centering
\includegraphics[width=0.9\linewidth]{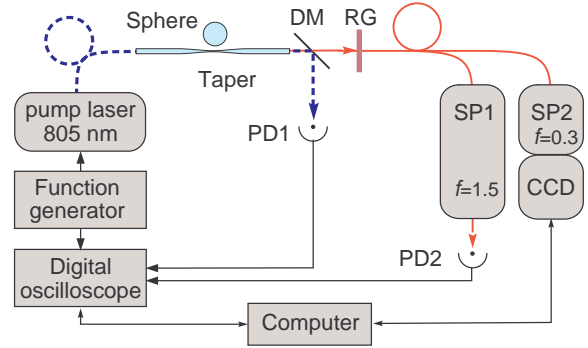}
\caption{(Color online)~Sketch of the experimental setup.  Pump laser and microlaser beams are shown in dotted blue and solid red, respectively. DM: dichroic
mirror;  RG:Schott filter RG850; SP1\&SP2 : spectrometers.}\label{setup}
\end{figure}

In our experiment, the microlaser is optically pumped by a laser injected in a WGM. In contrast to usual techniques, its frequency is not fixed, but
is continuously swept across the resonance, at a nearly constant power. The thermal effect converts this frequency modulation into a smooth sweep
of the power coupled into the cavity. The simultaneous monitoring of the injected power and the emitted signal thus directly provides a real-time measurement
of the pump--microlaser characteristic, conveniently displayed on an oscilloscope. This method is demonstrated on silica microspheres functionalized with neodymium, which reproducibly feature a sub-microwatt laser threshold. It is shown to be consistent with the results of a more conventional step-by-step method, relying on static thermal effect.

Our experimental setup is sketched in Fig.~\ref{setup}. 
The microsphere fabrication and the heart of the experiment where the microsphere is evanescently coupled to the fiber taper are explained in \cite{LinQian2010}. The taper is produced as described in \cite{OrucevicLefevre2007}.

The microspheres are functionalized by dip-coating in an alcoholic colloidal suspension of Nd-doped Gd$_2$O$_3$  nanocrystals~\cite{BazziBrenier2005}, followed by heating for a few seconds in order to anneal the 
emitters and to embed them into the sphere. The functionalized microspheres feature $Q$-factors ranging from $10^7$ to $10^8$ at pump wavelength 
$\lambda_p\approx805\,\mathrm{nm}$ and at least $10^8$ at emission wavelength 
$\lambda_e \approx 1083\,\mathrm{nm}$. 
The corresponding Nd content in the WGM volume is below 10~ppm, two orders of magnitude smaller than in \cite{SandoghdarTreussart1996}.
Since aging problems due to water deposition are often observed, the  typical $66\,\mathrm{\mu m}$ diameter silica microsphere
described here was stored for a few days in order to achieve stable lasing conditions. Its $Q$-factor was reduced to $8.5\times10^6$ at
$\lambda_p$ and $4.7\times10^7$ at $\lambda_e$. These very high figures lead to low lasing threshold and an even lower thermal
bistability threshold.

The pump is a free running laser diode at $\lambda_p\simeq805\,\mathrm{nm}$, coupled to the fiber taper and tuned to a narrow linewidth and low
angular order WGM, selected by the method of \cite{LinQian2010}. At the taper output, the transmitted pump and the emitted light coupled out by the
taper are separated by a dichroic mirror. The transmitted pump is measured by the silicon detector PD1. The emitted light is either filtered using
a high resolution ($\sim0.01\,\mathrm{nm}$) spectrometer (SP1) and measured with PD2, or analyzed at a lower resolution with a short focal length
spectrometer (SP2) equipped with a spectroscopic camera (CCD), providing broadband emission spectra of Nd$^{3+}$ ions.

\begin{figure}
\centering
\includegraphics[width=0.9\linewidth]{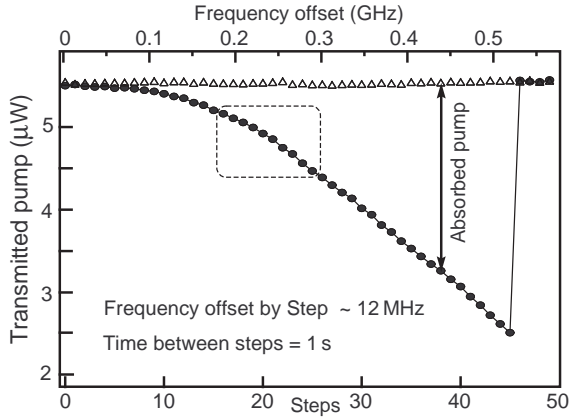}
\caption{\small Transmitted pump vs steps. Empty triangles: reference obtained for a large gap.  Full circles: dip observed for a gap of
$\sim150\,\mathrm{nm}$. The vertical arrow denotes the ``absorbed pump power''. The dotted zone defines the spectra plotted in
Fig.~\ref{f:laserspeccarac}.}
\label{f:trigpump}
\end{figure}

\begin{figure}
\centering \includegraphics[width=0.9\linewidth]{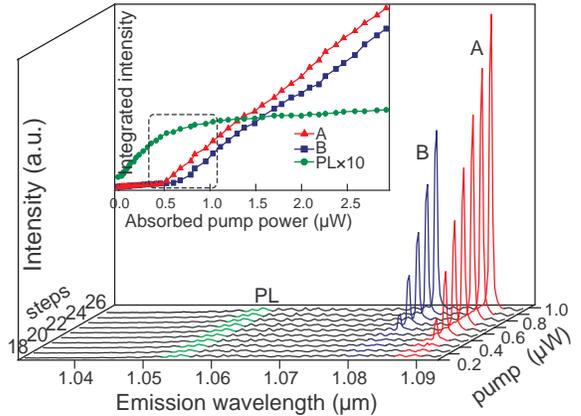}
\caption{\small (Color online) Emission spectra with increasing absorbed pump power for a step range
 denoted by the dashed rectangle in inset and Fig.~\ref{f:trigpump}. Inset: area of peaks A and B and of PL.
 }\label{f:laserspeccarac}
\end{figure}

To establish a firm evidence of laser action and of the threshold value, we present first  a step-by-step variant of our method. In this
method, the frequency of the pump laser diode is decreased stepwise, at a time rate of about 1 point every $1\,\mathrm{s}$. This time is much longer
than the time constants of thermal effect ($\lesssim 15\,\mathrm{ms}$) and of laser dynamics, such that each step corresponds to a steady-state
operation, and to the static thermal bistability condition. In Fig.~\ref{f:trigpump}, we plot the transmitted pump intensity (PD1) as a function of
the step number (ie pump frequency). The \emph{absorbed} pump power is given by the absorption dip, obtained by difference between the nearly
constant reference measured with uncoupled sphere and the loaded transmission, as shown by the vertical arrow.

Some emission spectra recorded on CCD during this frequency-steeping  are plotted in 3-D in Fig.~\ref{f:laserspeccarac}. They clearly evidence a
bi-mode laser operation with a threshold of about $500\,\mathrm{nW}$. This is confirmed by the inset, where the area of the lasing peaks A and B,
and of the $1054\,\mathrm{nm}$ photoluminescence (PL), are plotted as a function of the absorbed pump power. The PL clearly exhibits a
quasi-saturation at threshold, an expected signature of population clamping.

\begin{figure}
\centering
\includegraphics[width=0.9\linewidth]{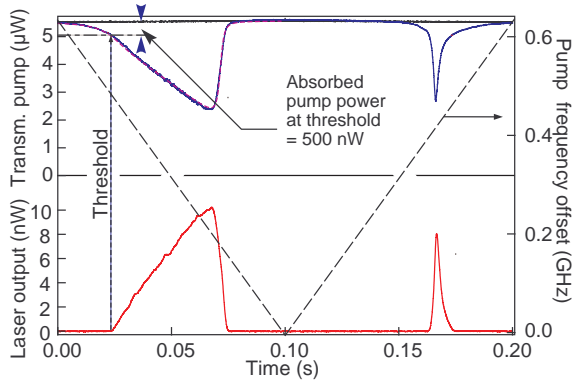}
\caption{\small (Color online) Simultaneous recording of the absorption dip and microsphere emission. Upper curves correspond to transmitted pump
power. Black: reference transmission; Blue: absorption dip; Purple dotted:  best fit to the theoretical transmission curve deduced from
Eq.~(\ref{e:thermalequation}). Lower red: measured laser emission. Dashed black: pump laser frequency offset.}\label{f:thermal}
\end{figure}

We now move on to the real-time method. The pump laser is then continuously swept across the WGM resonance, over the same frequency range of 
$\sim 0.6\,\mathrm{GHz}$ and the incident pump power is kept lower than $10\,\mathrm{\mu W}$ in the whole experiment. We now use SP1 to filter out a
\emph{single} emission WGM (peak A in Fig.~\ref{f:laserspeccarac}).  The transmitted pump intensity measured on PD1 and the filtered laser signal measured on PD2 are simultaneously
monitored on a digital oscilloscope, leading to the curves plotted in Fig.~\ref{f:thermal}. In this figure, the pump signal (upper blue curve)
exhibits a strong thermal dynamical bistability, characterized by the asymmetry between the decreasing and increasing frequency half-periods. The
purple dotted curve is a best fit to the numerical solution of the differential equation for the temperature difference $\theta(t)=T(t)-T_0$\,:
\begin{equation}\label{e:thermalequation}
\frac{\mathrm{d}\theta}{\mathrm{d}t} = \frac{A\,P_{\mbox{\tiny inc}}}{\Delta^2(t)+(\gamma/2)^2} - \gamma_{th}\, \theta(t) \ ,
\end{equation}
where the first term, accounting for the heating, is proportional (with a constant $A$) to the incident
power $P_{\mbox{\tiny inc}}$ and to a Lorentzian-shaped curve involving the effective detuning $\Delta(t)=\omega(t)-\omega_0-K\theta(t)$, where
$\omega(t)$ is the  pump frequency, and $K$ the cavity thermo-optic coefficient. The second term represents the thermal leak. As previously, the baseline (black  curve) was recorded with a large coupling gap, 
and the absorbed pump power is determined by difference with the baseline.

\begin{figure}
\centering \includegraphics[width=0.9\linewidth]{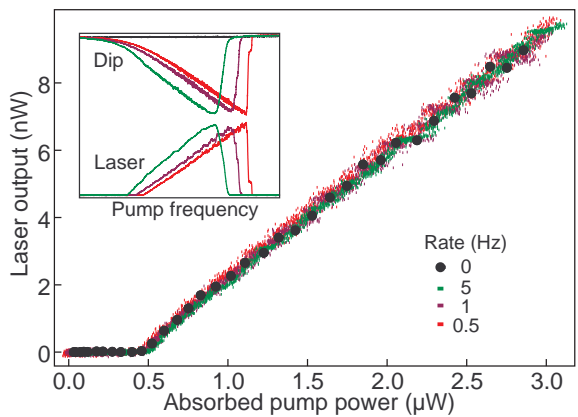}
\caption{\small (Color online) Light-light curve, for different scan speeds. Inset: corresponding transmitted pump and
laser output
curves.}\label{f:ThermalD}
\end{figure}

One notices that the onset of microsphere emission occurs later than the start of  pump absorption, demonstrating a threshold effect.
An X-Y plot, as used in Fig.~\ref{f:ThermalD}, allows to measure the threshold and estimate the differential efficiency  in \emph{real-time}. This
method is much quicker and easier to set-up than the standard power sweep and enables an efficient optimization in real-time.

In order to demonstrate the consistency of this method, we swept the pump laser at different repetition frequencies between $0.5\,\mathrm{Hz}$ and
$25\,\mathrm{Hz}$. The modulation frequency affects  the resonance shape, but for slow enough frequencies, the X-Y plot  provides the same
light--light characteristic, as shown in Fig.~\ref{f:ThermalD}. In this figure the characteristics obtained for modulations frequencies of  $0.5$, 1
and $5\,\mathrm{Hz}$ are plotted with red, purple, and green dots, respectively. The very good overlap of these curves demonstrates that the
observed threshold and slope efficiency do not depend on the repetition rate. This is no longer true when using higher frequencies (typically above
$10\,\mathrm{Hz}$), where an increasing departure (not shown) is observed, mostly due to the weakening of the thermal effect, as we have verified
thanks to a numerical simulation based on (1). To fully demonstrate that the observed light-light curve is the same as it
would be observed in steady state, we have combined the real-time method with the frequency-steeping described above. By recording at each step the
absorbed pump power and the microlaser power transmitted by SP1 on PD2, we have obtained the curve displayed with black circles in
Fig.~\ref{f:ThermalD}, in perfect agreement with the real-time characteristic. It is worth mentioning that this method uses non-steady state thermal
effect, while the much faster laser dynamics is nearly in steady-state and can adiabatically follow the temperature and pump power variations.

In summary, we have developed a simple method to characterize the laser characteristic of rare-earth based microlasers, taking advantage of the
very high quality factor of the doped microcavities. This results in a strong thermal effect, allowing to control the pump power by the mean of
frequency sweeping. This method can work at a rate as high as $10\,\mathrm{Hz}$, allowing a real-time optimization of the coupling
conditions. This approach allows to demonstrate a sub-microwatt threshold neodymium-based laser, and will be applied for further optimizations to
be published in a near future.

This work has been supported by the Region Île-de-France in the framework of C'Nano IdF (Nanoscience competence center of Paris Region). G.L.
acknowledges support from the China Scholarship Council.

\end{document}